# Cosmology with the Weyl-Dirac approach

**Mark Israelit**

Department of Physics and Mathematics, University of Haifa – Oranim, 36006 ISRAEL.
E-mail Israelit@macam.ac.il

Some problems of cosmology: the big bang singularity, the origin of conventional matter, of dark matter and of dark energy may be successfully described and treated in the framework of the Weyl-Dirac theory. This theory, being a minimal expansion of Einstein's general theory of relativity, contains in addition to the metric tensor $g_{\mu\nu} = g_{\nu\mu}$ the Weyl connection vector $w^{\mu}$ and the Dirac gauge function $\beta$. From these geometrically based quantities one obtains the behavior of our universe. The Weyl connection vector $w^{\mu}$ existing in microcells creates dark-matter particles, weylons. In the very early universe $\beta$ creates matter, whereas in the present dust period $\beta$ forms dark energy, the latter causing cosmic acceleration. In the present dust dominated period, around a massive body the $\beta$-dark energy forms a ball-like concentration having negative mass and negative pressure. These $\beta$-balls may cause an additional acceleration of the expanding universe.
The Weyl-Dirac theory is a classical (non-quantum) geometrically based framework appropriate for describing and searching cosmology.

## 1. Introduction

During the XIX and XX centuries Cosmology became an exact science mainly due to the theoretical basis of Einstein's general theory of relativity (GR) (Einstein, 1916, 1956) and due to the rapid development of optical and electronic devices. However even today it suffers from a number of problems (Weinberg, 2008), (Durer & Marteens, 2008), (Kamionkowski, 2007), (Ratra & Vogeley, 2008) that are in the spotlight of discussions and research. Amongst these problems are the following:

1. The Big Bang singularity of St. Weinberg's model (Weinberg, 1972) and connected difficulties, such as the flatness problem, the homogeneity and isotropy problem and the horizon problem (Guth, 1981). To overcome them complicated and theoretically insufficiently based scenarios were introduced, e.g. the inflationary scenario (Linde 1982, 1984, 2005).

2. The origin of conventional matter filling the presently observed universe is another unsolved mystery in cosmology ( Lima, Germano, Abramo, 1996). The question is: what brought matter into being?

3. Dark matter (DM) was postulated by Fritz Zwicky (Zwicky, 1933) already in 1933 and became very popular in the last 25 years. However the nature and origin of DM is under discussion still.

4. In nineties of the XX century a new phenomenon was discovered, - the present cosmic acceleration (Perlmutter et al, 1998, 1999), (Riess et al, 1998), (Garnavich et al, 1998) of our expanding universe. The factor causing this cosmic acceleration, dark energy (DE), is now widely discussed, however its nature and origin is unknown still. Thus, what is DE? More is unknown than is known. We know how much dark energy there is because we know how it affects the expansion of the Universe. Other than that, it is a complete mystery. But it is an important mystery. It turns out that roughly 70% of the Universe is dark energy; dark matter makes up about 25%; the rest: all "normal" matter, all stars, intergalactic gas etc. make less than 5% of the Universe.

Concerning the Big Bang problem, it would be interesting to remind that eighty years ago George Lemaître (Lemaitre, 1933) introduced the term Phoenix Universe to describe an oscillatory universe. However, during ~50 years the big bang model remained the popular one.

Twenty four years ago in the framework of Einstein's General Theory of Relativity a singularity-free oscillating cosmological model was proposed (Israelit & Rosen, 1989, 1993), (Rosen & Israelit, 1991). In that model the spatially closed universe began its present expansion phase from a cosmic egg filled by prematter, having the huge Planckian density, $\rho_{PL} = 3.83 \times 10^{65} \, cm^{-2}$, and satisfying the prematter equation of state, $P = -\rho$. The universe achieves the maximum radius and then undergoes a contraction phase going back to the cosmic egg. The period of oscillations of that model is $\sim 1.2 \times 10^{12} \, yr$. This singularity-free model was further developed by F. Cooperstock and collaborators (Starkovich & Cooperstock, 1992), (Bayin, Cooperstock & Faraoni, 1994). In these papers the prematter was attributed to a scalar field obeying a Klein-Gordon equation.

Recently an interesting revived phoenix model was proposed (Lehners Steinhardt & Turok, 2009), (Lehners & Steinhardt, 2009). This new cyclic model of the universe incorporates dark energy and cosmic acceleration in an essential way. The theory makes use of string theory and M-theory in which space-time consists of two brane-worlds separated by a tiny gap along an additional spatial dimension. One of these brane-worlds is our Universe. The big bang corresponds to a collision between the brane worlds, followed by a rebound. Matter, space and time exist before as well as after, and it is the events that occur before each bang that determine the evolution.

Let us turn to the problem of matter creation. What brought matter into being? There are many works dealing with this problem from various standpoints. There are works, where the matter was created by scalar fields (Starkovich & Cooperstock, 1992), (Bayin, Cooperstock & Faraoni, 1994), by electromagnetic geons (Perry & Cooperstock, 1999). Very interesting are the works by Heinz Dehnen et al (Dehnen & Ghaboussi, 1986), (Dehnen, Ghaboussi & Israelit, 1987), (Bezares-Roder & Dehnen, 2007), in which microscopic Yang-Mills and Higgs fields induce matter and the gravitational field of Einstein's general theory of relativity. During the last decades became popular Wesson's Induced Matter Theory (IMT), known also as the Space Time Matter Theory, which is based on the Kaluza-Klein approach. Regarding our 4-dimensional universe as a hyper-surface embedded in a 5-dimensional "bulk", Wesson et al (Wesson, 1992a, 1992b, 1999), (Overduin & Wesson, 1997), (Halpern & Wesson, 2006) have shown that the matter in our universe is induced by the geometry of the bulk, the latter being warped, but empty from matter.

Recently it was shown, that in Wesson's IMT the geometry of the 5-dimensional bulk is rather a Weylian one, than a Riemannian. The Weyl-Dirac version of Wesson's IMT was presented (Israelit, 2005a, 2005b). In that version classical models of fundamental particles induced by the bulk in our 4-dimensional universe were proposed (Israelit, 2007, 2008). It was also shown (Israelit, 2009) that conventional matter, dark matter and cosmic acceleration may be induced in the 4-universe by the Weyl-Dirac 5-bulk.

The standpoint of the present writer is that cosmological problems have to be considered and solved in a framework that is as close as possible to Einstein's General Theory of Relativity without adding fields, which are not geometrically based.

**In the present work, it is shown that the 4-dimensional Weyl-Dirac (W-D) theory, which is a minimal expansion of Einstein's theory, satisfies the above requirements and is a suitable framework for searching cosmological problems. In this framework, conventional matter, dark matter as well dark energy is created by geometry.**

**2. Basic ideas of the Weyl-Dirac theory**

After Einstein (Einstein, 1916, 1956 ) developed his general theory of relativity, in which gravitation is described in terms of geometry, Weyl (Weyl, 1918, 1919, 1923) proposed a more general theory, in which electromagnetism is also described geometrically. However, this theory had some unsatisfactory features and did not gain general acceptance. Later Dirac (Dirac, 1973) returned to Weyl's theory but introduced modifications which removed the earlier difficulties. Nathan Rosen (Rosen, 1982) in discussing the Weyl-Dirac (W-D) theory pointed out that Dirac had chosen a particular value for a parameter appearing in the W-D variational principle and that this parameter could be taken differently. Rosen showed that the parameter could be chosen so that, instead of an electromagnetic field, one gets a Proca (Proca, 1936) vector field, which may be interpreted as an ensemble of particles having finite mass and spin 1. It was suggested that these particles named weylons might constitute most of the dark matter (Israelit & Rosen, 1992, 1994). The W-D framework contains three geometric quantities: the metric tensor $g_{\mu\nu} = g_{\nu\mu}$, the Weyl connection vector $w^\mu$ and the Dirac gauge function $\beta$; from these geometrically based quantities one obtains the behavior of our universe.

Below a concise description of the Weyl-Dirac framework is given. Weyl issued from the dominance of light rays for physical measurements. Accordingly, he regarded the light cone as the principal phenomenon describing the space-time. This idea brought Weyl to regard rather the isotropic interval $ds^2 = 0$ as invariant, than an arbitrary line-element $ds^2 = g_{\mu\nu} dy^\mu dy^\nu$ between two space-time events. In the Weyl geometry, the metric interval between two events as well the length of a given vector is no more constant, it depends on an arbitrary multiplier, the gauge function. In the process of parallel displacement both, the direction and the length of a vector change.

If a vector having gauge-invariant contravariant components $B^\mu$ is displaced by $dx^\nu$, it changes by

$$dB^\mu = -B^\sigma \Gamma^\mu_{\sigma\nu} dx^\nu \qquad (1)$$

where $\Gamma^\lambda_{\mu\nu}$ is a connection. By the same displacement the length $B = (B^\mu B^\nu g_{\mu\nu})^{\frac{1}{2}}$ is changed by

$$dB = B w_\nu dx^\nu \qquad (2)$$

In (2) $w_\nu$ is the Weyl length connection vector. In Weyl's geometry, in addition to the usual coordinate transformations (CT) appear also Weyl gauge transformations (WGT), e.g. $g_{\mu\nu} \to \widehat{g}_{\mu\nu} = e^{2\lambda} g_{\mu\nu}$; $g^{\mu\nu} \to \widehat{g}^{\mu\nu} = e^{-2\lambda} g^{\mu\nu}$ where $\lambda(x^\nu)$ is a differentiable function; generally, if $\widehat{\Psi}^{\sigma\ldots}_{\mu\nu\ldots} = e^{n\lambda}\Psi^{\sigma\ldots}_{\mu\nu\ldots}$, the quantity $\Psi^{\sigma\ldots}_{\mu\nu\ldots}$ is called a gauge covariant of weight **n**, and if $n = 0$, the quantity is gauge invariant.

For relation (2) to hold in any gauge, one must have the following WGT for $w_\mu$:

$$w_\mu \to \widehat{w}_\mu = w_\mu + \lambda_{,\mu} \qquad (3)$$

(A comma denotes partial differentiation)

The connection $\Gamma^\lambda_{\mu\nu} = \Gamma^\lambda_{\nu\mu}$ is expressed by the metric tensor $g_{\mu\nu}$ and the Weyl length connection vector $w_\nu$ as

$$\Gamma^\lambda_{\mu\nu} = \{^\lambda_{\mu\nu}\} + g_{\mu\nu} w^\lambda - \delta^\lambda_\nu w_\mu - \delta^\lambda_\mu w_\nu \qquad (4)$$

with $\{^\lambda_{\mu\nu}\}$ being the conventional Christoffel symbol formed from $g_{\mu\nu}$.

Now, let a vector $B^\mu$ be transported by parallel displacement around an infinitesimal closed parallelogram having sites $dx^\mu, \delta x^\nu$. Then one find for the total change of the component

$$\Delta B^\lambda = B^\sigma K^\lambda_{\sigma\mu\nu} dx^\mu \delta x^\nu \qquad (5)$$

While the length of the vector is changed by

$$\Delta B = B W_{\mu\nu} dx^\mu \delta x^\nu \qquad (5a)$$

Here $K^\lambda_{\sigma\mu\nu} = -\Gamma^\lambda_{\sigma\mu,\nu} + \Gamma^\lambda_{\sigma\nu,\mu} - \Gamma^\alpha_{\sigma\mu}\Gamma^\lambda_{\alpha\nu} + \Gamma^\alpha_{\sigma\nu}\Gamma^\lambda_{\alpha\mu}$, is the curvature tensor formed of the connection $\Gamma^\lambda_{\mu\nu}$, and the Weyl length curvature tensor $W_{\mu\nu}$ is given by

$$W_{\mu\nu} = w_{\mu,\nu} - w_{\nu,\mu} \qquad (6)$$

EQ. (3) and (6) led Weyl to identify $w_\mu$ with the potential vector, and $W_{\mu\nu}$ with the strength tensor of the electromagnetic field. In order to get a geometrically based description of gravitation and electromagnetism, Weyl derived his field equations from a variational principle $\delta I = 0$ with the action $I = \int L(-g)^{\frac{1}{2}} d^4 x$. The Lagrangian density $L$ was built of the curvature tensors $K^\lambda_{\sigma\mu\nu}$, and $W_{\mu\nu}$. Weyl had to take an action, which was invariant under both coordinate transformations and WGT. For this he was forced to take an expression, which involved the square of Riemann's curvature scalar. This led to unsatisfactory equations for the gravitational field.

Dirac (Dirac 1973) presented a revised version of Weyl's theory. He introduced a positive scalar field $\beta(x^\nu) > 0$ of weight -1, which under WGT changes as

$$\beta \to \widehat{\beta} = e^{-\lambda}\beta \qquad (7)$$

As the scalar $\beta$ defines uniquely the gauge, it is called the Dirac gauge function. With $\beta$, and making use of $K^\lambda_{\sigma\mu\nu}$ and $W_{\mu\nu}$, Dirac wrote the action integral as

$$I = \int \left[ W^{\lambda\rho}W_{\lambda\rho} - \beta^2 R + \sigma\beta^2 w^\lambda w_\lambda + 2\sigma\beta w^\lambda \beta_{,\lambda} + (\sigma+6)\beta_{,\rho}\beta_{,\lambda} g^{\lambda\rho} + \right. \tag{8}$$
$$\left. + 2\Lambda\beta^4 + L_{matter} \right] \sqrt{-g}\ d^4x$$

In (8) $\sigma$ is the Dirac parameter, $\Lambda$ is the cosmological constant, $R$ is the Riemannian curvature scalar and $L_{matter}$ is the Lagrangian density of matter. Varying in (8) $g_{\mu\nu}$, $w_\nu$, $\beta$ and choosing $\sigma = 0$, was obtained (Dirac, 1973) a geometrically based theory of gravitation and electromagnetism that in the Einstein gauge, $\beta = 1$, results in Einstein's general relativity theory and Maxwell's electrodynamics.

Nathan Rosen (Rosen, 1982) analyzing the W-D theory, showed that if one instead $\sigma = 0$ takes $\sigma < 0$ one gets a Proca vector field (Proca, 1936), which from the quantum mechanical point of view may be treated as an ensemble of massive particles having spin 1. In the following we will make use of the W-D theory enriched by Rosen's approach, where $\sigma < 0$.

The basic ideas of the Weyl-Dirac theory may be found in works of the founders, (Weyl, 1918, 1919, 1923), (Dirac, 1973). Discussions and detailed descriptions as well as extensions of the W-D theory are presented in Rosen's paper (Rosen, 1982), in a monograph (Israelit, 1999), in (Israelit, 2011) as well in an interesting paper by R. Carroll (Carroll, 2007).

## 3. The field equations

The field equations will be derived from the action integral (8) that is coordinate invariant and gauge invariant. Thus, the EQ-s will be both, coordinate covariant and gauge covariant. (Cf. (Israelit, 2011)). We *emphasize this double covariance*.

In a coordinate covariant framework we can choose a coordinate system of reference suitable for our observations. So, astronomical observations led us to the cosmological principle and the last in turn directed us to the F-L-R-W line element. Another example is the mass of an insular system that has meaning only in asymptotically flat systems of reference.

In a gauge covariant framework we can choose a certain $\beta$, that defines the gauge. So, with $\beta = 1$ we get Einstein's gauge, which in the W-D theory leads to GR and Maxwell's electrodynamics. More choices are possible and will be used below to describe cosmology.

Let us consider (8). Here conventional matter is presented by the Lagrangian $L_{matter}$ that generally may depend on all three variables, $g_{\mu\nu}$; $w_\nu$; $\beta$, and on an additional variable $\varphi$, this latter appearing only in $L_{matter}$, so that $\dfrac{\delta(L_{matter}\sqrt{-g})}{\delta\varphi} = 0$. Thus,

$$\delta(L_{matter}\sqrt{-g}) = 8\pi T^{\mu\nu}\sqrt{-g}\ \delta g_{\mu\nu} + 16\pi J^\lambda \sqrt{-g}\ \delta w_\lambda + \Psi\sqrt{-g}\ \delta\beta. \tag{9}$$

Here $T^{\mu\nu}$ is the momentum-energy tensor of conventional matter, $J^\lambda$ is the current of the Weyl field $w_\lambda$ and $\Psi$ stands for the charge of the Dirac gauge scalar field $\beta$.

Varying in the action (8) the metric tensor, $g_{\mu\nu}$ we obtain the gravitational equation

$$G^{\mu\nu} = -\frac{8\pi}{\beta^2}\left(T^{\mu\nu} + M^{\mu\nu}\right) + \frac{1}{\beta^2}\left(4\beta^\mu\beta^\nu - g^{\mu\nu}\beta^\lambda\beta_\lambda\right) + \frac{2}{\beta}\left(g^{\mu\nu}\beta^\lambda_{;\lambda} - \beta^\nu_{;\lambda}g^{\lambda\mu}\right) +$$
$$+ \frac{\sigma}{\beta^2}\left(\beta^\mu\beta^\nu - \frac{g^{\mu\nu}}{2}\beta^\lambda\beta_\lambda\right) + \frac{\sigma}{\beta}\left(\beta^\mu w^\nu + \beta^\nu w^\mu - g^{\mu\nu}\beta_\lambda w^\lambda\right) +$$
$$+ \sigma\left(w^\mu w^\nu - \frac{g^{\mu\nu}}{2}w^\lambda w_\lambda\right) - g^{\mu\nu}\beta^2\Lambda \qquad (10)$$

In (10), $G^{\mu\nu}$ is the Einstein tensor of general relativity, the expression $4\pi M^{\mu\nu} = \left[\frac{1}{4}g^{\mu\nu}W^{\lambda\rho}W_{\lambda\rho} - W^\mu_{\ \lambda}W^{\nu\lambda}\right]$ stands for the momentum-energy tensor of the $w_\nu$ – field, while $\beta_\nu \equiv \beta_{,\nu} \equiv \frac{\partial\beta}{\partial x^\nu}$; $\beta^\mu \equiv g^{\mu\nu}\beta_\nu$ and $\beta^\mu_{;\nu} \equiv \frac{\partial\beta^\mu}{\partial x^\nu} + \{^\mu_{\sigma\nu}\}\beta^\sigma$. Setting in (10) $\sigma = 0$ and $\beta = 1$ one obtains the gravitational equation for G.R. and Maxwell's electrodynamics, whilst setting $w_\lambda = 0$; $\beta = 1$ leads to the EFE of G.R.

Varying in (8) the Weyl length connection vector, $w_\mu$, yields

$$4W^{\mu\nu}_{\ ;\nu} = 16\pi J^\mu + 2\sigma\beta\left(\beta^\mu + \beta w^\mu\right) \qquad (11)$$

We assume that conventional matter interact with Weylian matter only geometrically, so that in (9) the Weylian-current and $\beta$ – charge vanish

$$J^\mu = 0 \ ; \ \Psi = 0 \qquad (11a)$$

and EQ. (11) may be written as

$$W^{\mu\nu}_{\ ;\nu} = \frac{1}{2}\sigma\beta\left(\beta^\mu + \beta w^\mu\right) \qquad (11b)$$

This leads to

$$\left(\beta\beta^\mu + \beta^2 w^\mu\right)_{;\mu} = 0 \qquad (11c)$$

Finally, varying in (8) the gauge function $\beta$, one is led to the EQ.

$$R = \sigma\left(w^\lambda w_\lambda - w^\lambda_{;\lambda}\right) - (\sigma + 6)\frac{\beta^\lambda_{;\lambda}}{\beta} + 4\beta^2\Lambda + \frac{\Psi}{2\beta} \qquad (12)$$

With (11a) the contraction of EQ. (10) is identical with (12), so that Dirac's gauge function $\beta$ may be chosen arbitrarily; this is obviously a consequence of being our framework a gauge covariant one. Taking into account the almost arbitrary choice of the gauge $(\beta > 0)$, we are now left with two EQ-s, (10) and (11b).

According to observations the universe is on a large scale homogeneous and isotropic, so that in global EQ-s no vector functions appear. One can think about the space-time of the universe having a chaotic Weylian microstructure (cf. (Israelit & Rosen, 1992, 1994)) but being on a large scale homogeneous and isotropic. In order to describe both, the microstructure and the global cosmic features, we will regard the Weyl vector as composed of two parts, $w_\nu = w_{\nu\,\text{glob}} + w_{\nu\,\text{loc}}$, with $w_{\nu\,\text{glob}}$ being a global gradient vector, so that globally no vector fields will be present. The vector $w_{\nu\,\text{loc.}}$ will represent the locally restricted chaotically oriented fields existing in micro-cells.

Let us first consider the global Weyl field. For it one has from (11b)

$$W^{\mu\nu}_{;\nu\,\text{glob}} = \frac{1}{2}\sigma\beta\left(\beta^\mu + \beta\, w^\mu_{\text{glob}}\right) \tag{11d}$$

with $W_{\mu\nu\,\text{glob}} \equiv w_{\mu,\nu\,\text{glob}} - w_{\nu,\mu\,\text{glob}}$  In order to get $W^{\mu\nu}_{;\nu\,\text{glob}} = 0$ in the homogenous and isotropic universe, we take

$$\beta_\lambda + \beta\, w_{\lambda\,\text{glob}} = 0; \Rightarrow w_{\lambda\,\text{glob}} = -\frac{\beta_\lambda}{\beta} \tag{13}$$

By condition (13), EQ. (11d) is satisfied identically, while EQ. (10) takes the form

$$G^{\mu\nu} = -\frac{8\pi}{\beta^2} T^{\mu\nu}_{\text{matter}} + \frac{1}{\beta^2}\left(4\beta^\mu\beta^\nu - g^{\mu\nu}\beta^\lambda\beta_\lambda\right) + \frac{2}{\beta}\left(g^{\mu\nu}\beta^\lambda_{;\lambda} - \beta^\nu_{;\lambda}g^{\lambda\mu}\right) - g^{\mu\nu}\beta^2\Lambda \tag{14}$$

EQ. (14) is the global gravitational equation, in it $T^{\mu\nu}_{\text{matter}}$ represents both, the ordinary matter as well the weylon dark matter.

### 4. The global field of the isotropic homogeneous universe

For cosmological considerations in the homogeneous, isotropic universe we are left with the gravitational EQ. (14). We recall that $T^{\mu\nu}_{\text{matter}}$ is the momentum-energy tensor of both, conventional and dark matter – weylons created in the micro cells. The closed homogeneous and isotropic universe may be described by the F-L-R-W line element

$$ds^2 = dt^2 - a^2(t)\left[\frac{dr^2}{1-r^2} + r^2 d\vartheta^2 + r^2\sin^2\vartheta\, d\varphi^2\right] \tag{15}$$

with $a(t)$ being the cosmic scale parameter. By the assumed symmetry of the universe Dirac's gauge function depends only on the cosmic time $t$.

Taking into account (15) and inserting into (14) $\beta(t)$ we obtain the EQ-s explicitly

$$G^0_0 \equiv -8\pi\,\rho_{\text{glob}} = -\frac{8\pi}{\beta^2}\rho_{\text{matter}} + 3\frac{(\dot\beta)^2}{\beta^2} + 6\frac{\dot a}{a}\frac{\dot\beta}{\beta} - \beta^2\Lambda \tag{16}$$

$$G^1_1 \equiv 8\pi\,P_{\text{glob}} = \frac{8\pi}{\beta^2}P_{\text{matter}} - \frac{(\dot\beta)^2}{\beta^2} + 2\frac{\ddot\beta}{\beta} + 4\frac{\dot a}{a}\frac{\dot\beta}{\beta} - \beta^2\Lambda \tag{17}$$

Here a dot denotes partial derivative with respect to $t$ and $\rho_{\text{matter}}$ and $P_{\text{matter}}$ stand for the density and pressure of both, conventional matter and weylon DM. From (16, 17) we obtain the cosmological EQ-s.

$$\frac{\dot a^2}{a^2} = \frac{8\pi}{3\beta^2}\rho_{\text{matter}} - \frac{(\dot\beta)^2}{\beta^2} - 2\frac{\dot a}{a}\frac{\dot\beta}{\beta} + \frac{1}{3}\beta^2\Lambda - \frac{1}{a^2} \tag{18}$$

$$\frac{\ddot a}{a} = -\frac{4\pi}{\beta^2}\left(P_{\text{matter}} + \frac{1}{3}\rho_{\text{matter}}\right) + \frac{(\dot\beta)^2}{\beta^2} - \frac{\dot a\dot\beta}{a\beta} - \frac{\ddot\beta}{\beta} + \frac{1}{3}\beta^2\Lambda \tag{19}$$

Making use of the Bianchi identity one has

$$\dot{\rho}_{glob} = -3\frac{\dot{a}}{a}\left(\rho_{glob} + P_{glob}\right) \quad (20)$$

and with (16, 17) one obtains an energy relation

$$\frac{8\pi}{\beta^2}\dot{\rho}_{matter} = -\frac{3\dot{a}}{a}\frac{8\pi}{\beta^2}\left(\rho_{matter} + P_{matter}\right) + 16\pi\,\rho_{matter}\frac{2\dot{\beta}}{\beta^3} + 6\left(\frac{\dot{\beta}\ddot{\beta}}{\beta^2} - \frac{(\dot{\beta})^3}{\beta^3} + \frac{\ddot{a}\dot{\beta}}{a\beta} + \frac{\dot{a}(\dot{\beta})^2}{a\beta^2}\right) - 2\dot{\beta}\beta\Lambda \quad (21)$$

In equations (17-21) Dirac's gauge function $\beta(t)$ is involved. This can be chosen nearly arbitrarily in the W-D framework; the only restriction is $\beta(t) > 0$.

4a. Choosing the gauge function.

For our purpose it is convenient to introduce the Dirac gauge function $\beta$ as consisting of two parts,

$$\beta = U_1 \beta_{create} + U_2 \beta_{late} \quad (22)$$

where $U_1$ and $U_2$ are step functions given by:

$$U_1 = \begin{cases} 1, & \text{for } a_0 < a(t) < A \\ \tfrac{1}{2}, & \text{for } a = A \\ 0, & \text{for } a(t) > A \end{cases} \quad \text{and} \quad U_2 = \begin{cases} 0, & \text{for } a(t) < A \\ \tfrac{1}{2}, & \text{for } a = A \\ 1, & \text{for } a(t) > A \end{cases} \quad (23)$$

Making use of present cosmological observations we can fix $A = 1.544 \times 10^{-3}$ cm (Israelit, 2011). In EQ. (23), $a(t)$ is the cosmic scale parameter and $a_0$ is that of the universe at the very beginning. Further, $\beta_{create}$ is the gauge function in the very early universe, while $\beta_{late}$ is acting for all times when $a(t) > A$.

Let us choose $\beta_{create}(t)$ according to the following scenario. We suppose that the universe began its present expansion phase from a sphere having radius $a_0$ identical with the Planckian length, i.e. $a_0 \equiv l_P = 1.616 \times 10^{-33}$ cm. This initial state was stationary and at the very beginning no matter (ordinary or weylon DM) was present. There was only the Dirac gauge function $\beta_{create}(t)$. We assume that $\beta_{create}(t)$ had a huge maximum at the beginning ($t = 0$), that it is an even function of time and that it is rapidly decreasing. In the very early universe, $\beta_{create}$ has to be a matter creating function. There are of course many functions satisfying the above requirements. We will adopt the following:

$$\beta_{create} = \frac{B}{\cosh^n(\gamma t)} \quad (24)$$

In (24), B (Capital Beta) and $\gamma$ are positive constants and $n$ is a positive integer. These constants will be fixed below.

The function $\beta_{late}$ acts from the moment, when the radius of the universe $a(t) > A$; The value of $\beta_{late}$ is very close to 1 and its derivatives are almost negligible up to the dust dominated period, where it will create dark energy, the latter being responsible for the acceleration of our universe at present.

Let us take the gauge function $\beta_{late}$ as
$$\beta_{late} = 1 + C_1 \tanh[\delta_1(X-1)] + C_2 \tanh[\delta_2(X^2-1)] \qquad (25)$$
Here $X(t) = \dfrac{a(t)}{a_N}$, and $a_N$ stands for the present value of the cosmic scale parameter. The parameters $C_1; C_2; \delta_1; \delta_2$ as fixed in accordance with cosmological observations (Carroll, 2001), (Spergel et al, 2003, 2007) are:
$C_1 = -10^{-2}; \; C_2 = 10^{-2}; \; \delta_1 = 1.22 \times 10^2; \; \delta_2 = 3.10 \times 10^1$ (Cf. below section 6a)

### 4b. The beginning.

Let us consider the situation at the very beginning, $t = 0$, when no matter was present $\rho_{matter} = 0; \; P_{matter} = 0$. We rewrite (18) as
$$\left(\frac{\dot a}{a} + \frac{\dot\beta_{create}}{\beta_{create}}\right)^2 = \frac{8\pi}{3\beta_{create}^2}\rho_{matter} + \frac{1}{3}\beta_{create}^2 \Lambda - \frac{1}{a^2} \qquad (18a)$$
Inserting (24) and supposing that at $t=0$ there was a stationary state, $\left(\dfrac{\dot a}{a}\right)_{t=0} = 0$, we obtain
$$\frac{1}{3}\mathrm{B}^2 \Lambda = \frac{1}{a_0^2} \qquad (26)$$
Here $a_0 \approx l_P = 1.616 \times 10^{-33}$ cm and $\Lambda \approx 2.074 \times 10^{-58}$ cm$^{-2}$ (cf. (Carroll, 2011)). So that
$$\mathrm{B} \equiv \beta_{create}(t=0) \approx 0.744 \times 10^{62} \qquad (27)$$
Going back to EQ. (18) we see that close to the beginning it may be written as $\dot a^2 = \dfrac{1}{3}\mathrm{B}^2 \Lambda\, a^2 - 1$, which is satisfied by
$$a(t) = a_0 \cosh\left(\frac{t}{a_0}\right) \qquad (28)$$

Let us turn to EQ. (19). For the very early universe it takes the form
$$\frac{\ddot a}{a} = -\frac{\ddot\beta}{\beta} + \frac{1}{3}\beta^2 \Lambda \qquad (29)$$
By (26, 28, 29) we are led to the condition
$$-\left(\frac{\ddot\beta}{\beta}\right)_0 \doteq n\gamma^2 \ll \frac{1}{3}\mathrm{B}^2\Lambda = \frac{1}{a_0^2} = 3.83 \times 10^{65} \; cm^{-2} \qquad (30)$$
One sees that at the very beginning in the universe acts a huge acceleration
$$\frac{\ddot a}{a} \approx \frac{1}{3}\mathrm{B}^2\Lambda = \frac{1}{a_0^2} = 3.83 \times 10^{65} \; cm^{-2} \qquad (30a)$$
This acceleration will cause a high speed expansion immediately after the beginning; the universe will enter the expansion phase.

4c. Matter creation in the early universe.

Very close to the beginning, Dirac's gauge function $\beta_{create}$ creates matter. Indeed, taking into account that $\rho_{matter} = 0$; $P_{matter} = 0$ one has from (21)

$$\frac{8\pi}{\beta^2}\dot{\rho}_{matter} = 6\left(\frac{\dot{\beta}\ddot{\beta}}{\beta^2} - \frac{(\dot{\beta})^3}{\beta^3} + \frac{\ddot{a}\dot{\beta}}{a\beta} + \frac{\dot{a}(\dot{\beta})^2}{a\beta^2}\right) - 2\dot{\beta}\beta\Lambda \tag{31}$$

Expanding $\beta_{create}(t)$; as well $\frac{\dot{a}}{a}$; $\frac{\ddot{a}}{a}$ in time, one obtains from (31) for $\gamma t \ll 1$ the following matter creation relation

$$\dot{\rho}_{matter} = \frac{3}{4\pi} B^2 n^2 \gamma^4 t; \tag{32}$$

Thus, matter created during a very small time, $t$, after the beginning has the density

$$\rho_{matter}(t) = \frac{3}{8\pi} B^2 n^2 \gamma^4 t^2 \tag{32a}$$

We adopt the idea that there exists a limit value for $\rho_{matter}$ -- the Planck density[1] $\rho_{Pl} = 3.83 \times 10^{65} \text{cm}^{-2}$, so that, after achieving this limit density, the growing of matter density will be stopped. The density $\rho_{matter}$ has a saturation point. The moment, when that saturation takes place is unknown. Let as assume that it occurs very close to the beginning, at $t \doteq 5a_0 = 8.08 \times 10^{-33} \text{cm}$, We denote this moment as $t_{Pl}$. According to (28) at $t_{Pl}$ the radius is $a_{Pl} = 1.2 \times 10^{-31} \text{cm}$. Actually, $t_{Pl}$ and $a_{Pl}$ describe the beginning of the matter universe with $\rho_{Pl}$. With $t_{Pl}$ and $\rho_{Pl}$ one obtains from (32a) $n\gamma^2 \approx 2.98 \times 10^3 \text{cm}^{-2}$ a value that satisfies condition (30). The value of the power $n$ is arbitrary. It turns out that $n = 1000$ is a good choice for the calculations, so that $\gamma = 1.726 \text{cm}^{-1}$. These values will be hereafter adopted.

After $t_{Pl}$, when the sphere is filled with matter, EQ. (31) is no more holding. With nonzero $P_{matter}$; $\rho_{matter}$ one obtains from (18, 19, 21) the following energy relation

$$\dot{\rho}_{matter} = -3\left(\frac{\dot{a}}{a} + \frac{\dot{\beta}}{\beta}\right)\left(\rho_{matter} + P_{matter}\right) + 4\frac{\dot{\beta}}{\beta}\rho_{matter} \tag{31a}$$

As we know nothing about the relation between pressure and density we turn to EQ-s (18, 19). Making use of (28) we obtain in the very early universe the following equation of state (EoS)

$$P_{matter} = -\rho_{matter} \tag{33}$$

But this is exactly the EoS for prematter, a substance introduced in (Israelit & Rosen, 1989). Let us assume that the prematter period lasts up to $t_1 = 65 a_0 = 1.05 \times 10^{-31}$ cm.

By (33) we have from (31a)

---

[1] We take $\rho_{Pl} = 3.83 \times 10^{65} \text{cm}^{-2}$ in geometric quantities. In conventional quantities one has $\rho_{Pl} = 5.16 \times 10^{93} \frac{g}{cm^3}$

$$\dot{\rho}_{\text{matter}} = 4\frac{\dot{\beta}_{\text{create}}}{\beta_{\text{create}}}\rho_{\text{matter}} \qquad (34)$$

For small values of *t* one obtains $\rho_{\text{matter}} = \rho_{Pl} \cdot e^{-4n\gamma^2 t}$, from which follows that during the whole prematter period, $t_{Pl} = 5a_0 \leq t_{\text{prematter}} \leq t_1 = 65a_0$, the matter density remains equal to $\rho_{Pl}$. One can prove that EQ. (28) is holding during the whole prematter period, so that at its end $t_1 = 1.05 \times 10^{-31} cm$, $a_1 = a(t_1) = 1.37 \times 10^{-5} cm$.

4d. The transition period

The prematter period is followed by a transition period, during which prematter turns gradually into radiation. This transition lasts from the end of the prematter period: $t_1$, $a_1 = a(t_1)$ with $\beta_{\text{create}} = B$, $\rho_{\text{matter}} = \rho_{Pl}$, up to $a_{\text{Rad}} \approx 100 cm$, when the universe is filled with radiation and ultrarelativistic particles. Hereafter we will drop the subscript $_{\text{matter}}$ that appeared in $\rho_{\text{matter}}$ and $P_{\text{matter}}$.

During the transition period both, $\beta_{\text{create}}(t)$ and $\rho$ are decreasing gradually. Accordingly, the EoS is changing from the prematter one, $P = -\rho$, to the radiation, $P = \tfrac{1}{3}\rho$. For the this period the following EoS will be taken

$$P = \frac{1}{3}\rho\left(1 - 4\frac{\rho}{\rho_{Pl}}\right) \qquad (35),$$

so that if $\rho$ is close to $\rho_{Pl}$, EQ. (35) turns into (33), while for $\rho \ll \rho_{Pl}$ it gives the radiation EoS

$$P = \frac{1}{3}\rho \qquad (36)$$

Let as go to the energy relation (31a). Making use of (35) and taking into account that during the whole transition period one has $\left|\frac{\dot{\beta}}{\beta}\right| \ll \frac{\dot{a}}{a}$, we can rewrite (31a) as

$$\dot{\rho} + 4\frac{\dot{a}}{a}\left(\rho - \frac{\rho^2}{\rho_{Pl}}\right) = 0 \qquad (31c)$$

Integrating one obtains for the matter density

$$\rho = \frac{A^4 \rho_{Pl}}{A^4 + a^4} \qquad (37)$$

Further, making use of (35) one has for the pressure

$$P = \frac{A^4 \rho_{Pl}\left(\frac{1}{3}\cdot a^4 - A^4\right)}{\left(A^4 + a^4\right)^2} \qquad (37a)$$

In (37, 37a) the constant of integration *A* will be chosen as the radius of the universe at a moment inside the transition period, so that for $a \ll A$ equation (33) is valid, while for $a \gg A$ (36) holds. The value of *A* can be linked to present cosmological observations (Israelit, 2011). It turns out that $A = 1.544 \times 10^{-3} cm$. Making use of (28) one has for the according moment $t_A = 69.7 a_0 \approx 1.127 \times 10^{-31}$ cm. At this state there is still $\rho = \rho_{Pl}$.

Going back to (18) and making use of (37) one can derive the relation *a(t)* in the transition period. So, for $a(t) \leq A$ we obtain

$$a(t) = a_0 \cosh\left(\frac{t}{a_0} + Const\right) \tag{38}$$

For $a > A = 1.544 \times 10^{-3}$ cm when $\beta_{create} = 0$; $\beta_{late} \approx 1$ one obtains (Israelit, 2011)

$$\left. \frac{1}{2}\sqrt{1+x^4} - \frac{1}{2}\ln\frac{\sqrt{1+x^4}+1}{x^2} \right|_1^x = \kappa(t - t_A) \tag{39}$$

In (39) $x(t) \equiv \frac{a(t)}{A}$ and $\kappa^2 = \frac{8\pi}{3}\rho_{Pl} = 3.2 \times 10^{66}$ cm$^{-2}$. With $A = 1.544 \times 10^{-3}$ cm and $t_A = 1.127 \times 10^{-31}$ cm one can calculate for any $a(t)$ the time **t**.

## 5. Local fields, weylons.

Let us turn to the weylons, which make the DM. We can consider the local Weyl field, $w^\mu_{loc.}$, existing in a microscopic cell, by procedures developed in (Israelit & Rosen, 1992, 1994)

Due to the smallness of the cell, the Dirac gauge function is actually constant and the Einstein gauge, $\beta = 1$, may be taken. Moreover it will be assumed that no ordinary matter is in the micro-cell,

Making use of (11a) and of the condition $\beta = 1$, we rewrite EQ. (11) in the cell as

$$W^{\mu\nu}_{loc.\ ;\nu} = \frac{1}{2}\sigma w^\mu_{loc.} \tag{40}$$

and with a new constant $\kappa$ given by $\sigma = -2\kappa^2$, (Cf. (Rosen, 1982)) we rewrite this as

$$W^{\mu\nu}_{loc.\ ;\nu} + \kappa^2 w^\mu_{loc.} = 0 \tag{40a}$$

This is the covariant form of the Proca EQ. for a vector boson field (Proca, 1936). From (40) follow immediately the condition

$$w^\mu_{loc.\ ;\mu} = 0 \tag{40b}$$

that may be interpreted as a conservation law for the Proca current, $\kappa^2 w^\mu_{loc.}$. Inside the cell no conventional matter is present and the curvature is negligible, so that making use of $W_{\mu\nu\ loc.} \equiv w_{\mu;\nu\ loc.} - w_{\nu;\mu\ loc.}$ one obtains from (40a)

$$g^{\lambda\nu} w^\mu_{loc\ ;\lambda;\nu} + \kappa^2 w^\mu_{loc.} = 0 \tag{41}$$

EQ. (41) describes a vector field, that from the quantum mechanical point of view is represented by bosons of spin 1 and mass $m_w$, the latter in conventional units being

$$m_w = \frac{\kappa \hbar}{c} \quad (42)$$

It must be emphasized that the local Weyl fields existing in the cells are chaotically oriented, so that the vector sum of the spins vanishes, and only the mass effect is observed. Following (Israelit & Rosen, 1992) we assume that these bosons, named **weylons**, can make up a considerable part of cold dark matter. These weylons are analogous to photons and gravitons but differ in that they are massive. It was shown in (Israelit & Rosen, 1994) that weylons, which interact with conventional (luminous) matter only through gravitation, may be regarded as a gas of particles in thermal equilibrium and for plausible values of mass $m_w$ they obey Boltzmann statistics. Further, weylons having mass $m_w > 10 \text{MeV}$ constitute a cold dark matter form that may be fitted into cosmological models. Supposedly these bosons were created in the very early universe, when prematter was transforming into radiation and ordinary ultrarelativistic particles.

We can consider the temperature history of our universe. Making use of two remarkable monographs (Weinberg, 1972) and (Landau & Lifshitz, 1959) in a previous paper (Israelit & Rosen, 1989) a formula linking between the temperature in the universe and values of the cosmic scale parameter was derived,

$$T = \left(\frac{\rho_{Pl}}{\sigma}\right)^{1/4} \frac{A a^7}{\left(A^4 + a^4\right)^2} \quad (43)$$

In (43) $\sigma = 6.24 \times 10^{-64} \text{cm}^{-2} (\text{K})^{-4}$ stands for the Stefan-Boltzmann constant, further $\left(\frac{\rho_{Pl}}{\sigma}\right)^{1/4} = 1.574 \times 10^{32} \text{K}$ and $A = 1.544 \times 10^{-3} \text{cm}$.

One find that at the beginning of the matter universe ($a_{Pl} = 1.2 \times 10^{-31} \text{cm}$) the universe was extremely cold $T_{Pl} = 2.70 \times 10^{-165} \text{K}$, at the end of the prematter period ($a_1 = 1.37 \times 10^{-5} \text{cm}$) one had $T_1 = 5.24 \times 10^{17} \text{K}$. The temperature continues to increase in the transition period and when $a^4 = 7A^4$ we get $T_{max} = 7.41 \times 10^{31} \text{K}$. From this point on $T$ decreased. At the beginning of the radiation period ($a_{Rad} = 100 \text{cm}$) there was $T_{Rad} = 2.43 \times 10^{27} \text{K}$.

As said above, weylons were created in the micro-cells while prematter was transforming into radiation and ultrarelativistic particles, and the temperature was close to its maximum value $T_{max} = 7.4 \times 10^{31} \text{K}$. It is plausible to assume that the creation process ended at $T_{Rad} = 2.43 \times 10^{27} \text{K}$, when prematter was completely transformed into radiation. At this moment both, the weylons and conventionally matter had the same temperature, $T_{Rad}$. Thereafter, as the weylons did not interact with conventionally (luminous) matter, they developed in time independently and their total number in the universe remained constant, i. e. $n_w a^3 = N_{w.total} = \text{const}$. During the radiation period, the temperature of conventionally matter, $T_{conv.} \propto a^{-1}$, and at the beginning of the dust period $a_{Dust} \approx 3.208 \times 10^{26} \text{cm}$ the temperature was $T_{Dust.conv.} = 7.575 \times 10^2 K$. On the other hand the weylons constitute a gas of heavy particles, $10^5 GeV \geq m \geq 10 MeV$ (Cf. (Israelit & Rosen 1992, 1994)) and their temperature $T_W$ decreases rapidly. At present the conventionally matter is at $T_{N.conv.} \approx (2.7 - 3.1)K$, whereas the temperature of the weylon gas, that depends on the weylon mass, is in the interval $10^{-17} K \leq (T_W)_N \leq 10^{-10} K$.

In the early universe, at the time the weylon DM was created, it had a density that was negligibly small compared to that of ordinary matter. Because of burning out of radiation, the fractional abundance of weylon dark matter became important at the time when galaxies were formed and it may be that weylon DM trough its gravitational interaction with conventionally matter played an important stimulating role in this process.

## 6. The dust dominated universe.

### 6a. The present state.
*The subscribe "N" (now) indicates values at present.*

In EQ. (25) was introduced the Dirac gauge function that acts from $A = 1.544 \times 10^{-3}$ cm up to $a_{max}$, $\beta_{late} = 1 + C_1 \tanh[\delta_1(X-1)] + C_2 \tanh[\delta_2(X^2-1)]$. In the dust dominated period this function will create DE, the latter causing the present cosmic acceleration. It is important to note that in the early universe up to the beginning of the dust dominated era there must be $\beta_{late} \approx 1$, while the derivatives must take insignificant values. The function $\beta_{late}$ depends on four parameters $C_1; C_2; \delta_1; \delta_2$, as well on the present value of the scale parameter $a_N$. These parameters may be fixed by present cosmological observations as well by the requirement that in the early universe $\beta_{late}(X \ll 1) \approx 1$, the latter leading to the condition

$$C_1 \tanh[\delta_1] + C_2 \tanh[\delta_2] = 0 \qquad (44)$$

From (25) one has at present $\beta_{late}(X=1) = 1$. Let us go back to EQ-s (18, 19).

Hereafter in the present section we will discard the subscript "$_{late}$". It is convenient to consider the gauge function as depending on the scale parameter $\beta(a(t))$. Denote

$$\frac{d\beta}{da} \equiv \beta,_a, \dot{\beta} = \beta,_a \dot{a}; \quad \ddot{\beta} = \beta,_{aa}(\dot{a})^2 + \beta,_a \ddot{a}, \quad b \equiv \ln\beta \quad, \ddot{\beta} = \beta,_{aa}(\dot{a})^2 + \beta,_a \ddot{a},$$

$$b \equiv \ln\beta, b,_a = \frac{\beta,_a}{\beta}; \quad b,_{aa} = \frac{\beta,_{aa}}{\beta} - (b,_a)^2. \qquad (45)$$

Then in the dust dominated period ($P \ll \rho_m$) EQ.-s (18, 19) take the form

$$\frac{\dot{a}^2}{a^2} = \frac{8\pi}{3\beta^2}\rho_m + \frac{1}{3}\beta^2\Lambda - \frac{1}{a^2} - \left(\frac{\dot{a}}{a}\right)^2\left[a^2(b,_a)^2 + 2ab,_a\right] \qquad (46)$$

$$\frac{\ddot{a}}{a} = -\frac{4\pi}{3\beta^2}\rho_m + \frac{1}{3}\beta^2\Lambda - \left(\frac{\dot{a}}{a}\right)^2\left[a^2 b,_{aa} + ab,_a\right] - \frac{\ddot{a}}{a}(ab,_a) \qquad (47)$$

In (46, 47), $\rho_m$ stands for the matter density of both, luminous and weylon DM..

From (46, 47) one has the following expressions for the density and pressure of DE.

$$\frac{8\pi}{3}\rho_{d.e.} = -\left(\frac{\dot{a}}{a}\right)^2 \left[a^2(b_{,a})^2 + 2ab_{,a}\right]; \quad 4\pi P_{d.e} = \left(\frac{\dot{a}}{a}\right)^2 \left[a^2 b_{,aa} + ab_{,a}\right] + \frac{\ddot{a}}{a}(ab_{,a}) \quad (48)$$

Considering the present state with $\beta_N \equiv \beta(a_N) = 1$, it is convenient to introduce the critic density $\rho_c \equiv \frac{3}{8\pi}H^2$, and the density parameters: $\Omega_m = \left(\frac{\rho_m}{\rho_c}\right)_N = \frac{8\pi}{3}\frac{(\rho_m)_N}{H^2}$; $\Omega_\Lambda = \frac{\Lambda}{3H^2}$; $\Omega_k = -\frac{1}{(a_N H)^2}$ and $\Omega_{DE} = \left(\frac{\rho_{DE}}{\rho_c}\right)_N = \frac{8\pi}{3}\frac{(\rho_{DE})_N}{H^2}$. One can also consider the pressure parameter of DE, $\Pi_{DE} = \left(\frac{P_{DE}}{\rho_c}\right)_N$. As the present value of radius of the universe $a_N$ is not measurable, we introduce a flatness parameter, $\eta = a_N H$. Large values of $\eta$ describe an almost flat universe, whereas small values – a well closed universe. We can rewrite (46) at present as

$$\Omega_{total} \equiv \Omega_m + \Omega_\Lambda + \Omega_{DE} = 1 + \frac{1}{\eta^2}; \quad (49)$$

One can make use of popular now data (Spergel, 2003, 2007): $H = 0.72 \times 10^{-28} \text{cm}^{-1}$; $\Lambda \approx 2.074 \times 10^{-58} \text{cm}^{-2}$; and the present overall $\rho_{tot} = 7.3488 \times 10^{-58} \text{cm}^{-2}$. Then one gets $\eta \approx 2.31$ and $a_N = \frac{\eta}{H} = 3.208 \times 10^{28}$ cm for the present value of the cosmic scale parameter. Further introducing the present acceleration parameter $Q_N \equiv \left(\frac{\ddot{a}}{a}\right)_N \frac{1}{H^2}$, we obtain from (47)

$$Q_N = -\frac{1}{2}\Omega_m + \Omega_\Lambda - (1+Q_N)a_N(b_{,a})_N - a_N{}^2(b_{,aa})_N \quad (50)$$

It was shown (Israelit, 2011) that fixing a fore-given $Q_N$ one can obtain the values of $C_1$; $C_2$; $\delta_1$; $\delta_2$. So, choosing $Q_N = 05$ and taking $a_N = 3.208 \times 10^{28}$ cm, one has the set

$$C_1 = -10^{-2}; \delta_1 = 1.22 \times 10^2; C_2 = 10^{-2}; \delta_2 = 3.1 \times 10^1. \quad (51)$$

These values are adopted in this work.

With (51) we have up to the dust period (for $a < 10^{-2} a_N$) $\beta_{late} \doteq 1$. Also in the early and late dust period $\beta_{late} \approx 1$. It slightly differs from 1 only in the interval $0.95 \leq X \leq 1.1$. However at present we have $\beta_{late}(X = 1) = 1$ exactly, (Cf. (25)).

6b. Dynamics in the dust dominated universe

The dust dominated period lasts from $0.01 a_N$ up to the maximum radius of the universe, $a_{max}$. To understand the behavior of physical quantities during this period, one can go back to (46 - 48). One can make use also of the energy relation in the dust dominated period, $\rho(a) = \dfrac{Const. \beta(a)}{a^3}$. As $\beta(a_N) = 1$, one has $Const. = \rho_N a_N^3$, so that

$$\rho(a) = \rho_N \beta(a) \left(\dfrac{a_N}{a}\right)^3. \tag{52}$$

Taking into account that during the whole dust period $\beta_{\text{late}} \approx 1$, we have from EQ. (46)

$$\left(\dfrac{\dot{a}}{a}\right)^2 (1 + ab,_a)^2 = \dfrac{\Lambda}{3} - \dfrac{1}{a^2} + \dfrac{8\pi}{3} \rho_N \left(\dfrac{a_N}{a}\right)^3 \tag{53}$$

That may be rewritten as

$$\left(\dfrac{\dot{a}}{a}\right)^2 (1 + ab,_a)^2 = \dfrac{\Lambda}{3} - \dfrac{1}{a_N^2 X^2} + \dfrac{8\pi}{3} (\rho_m)_N \dfrac{1}{X^3}. \tag{54}$$

From (54) follows that $\left(\dfrac{\dot{a}}{a}\right)^2$ has a minimum at $X_{\max} \approx 2.66$, so that $a_{\max} = 2.66 a_N$.

Let us introduce the current expansion parameter $h \equiv \dfrac{\dot{a}}{a}$ at any $a(\tau)$ as well the current density parameters:

$$\omega_m \equiv \dfrac{8\pi \rho_m}{3h^2} = \dfrac{8\pi}{3h^2} \rho_N \left(\dfrac{a_N}{a}\right)^3 = \Omega_m \left(\dfrac{H}{h}\right)^2 \left(\dfrac{a_N}{a}\right)^3, \quad \omega_\Lambda \equiv \dfrac{\Lambda}{3h^2} = \Omega_\Lambda \left(\dfrac{H}{h}\right)^2,$$

$$\omega_k \equiv -\dfrac{1}{h^2 a^2} = -\dfrac{1}{\eta^2} \left(\dfrac{H}{h}\right)^2 \left(\dfrac{a_N}{a}\right)^2 \text{ and } \omega_{DE} \equiv \dfrac{8\pi \rho_{DE.}}{3h^2} = -\left[a^2(b,_a)^2 + 2ab,_a\right]. \text{ We have}$$

also the current acceleration parameter, $Q \equiv \dfrac{\ddot{a}}{a} \dfrac{1}{h^2}$. Thus, we can rewrite (46, 47) as

$$\omega_m + \omega_\Lambda + \omega_k + \omega_{DE} = 1 \tag{46a}$$

and

$$Q(1 + ab,_a) = -\dfrac{1}{2} \omega_m + \omega_\Lambda - \left[a^2 b,_{aa} + ab,_a\right] \tag{47a}$$

From the definitions one has $\dfrac{\omega_\Lambda}{\omega_m} = \dfrac{\Omega_\Lambda}{\Omega_m} X^3$, $\left|\dfrac{\omega_k}{\omega_m}\right| = \dfrac{1}{\eta^2 \Omega_m} X$ and $\left|\dfrac{\omega_\Lambda}{\omega_k}\right| = \eta^2 \Omega_\Lambda X^2$, so that in the late universe, $X > 1$, the contributions of $\Lambda$ and of the spatial curvature are dominant, while in the early dust universe, $X < 1$, that of matter is dominant.

The density parameter of DE, $\omega_{DE}$ as well $\rho_{DE}$ both are negative and insignificant for $0.01 \leq X \leq 0.9$. In the interval $0.995 \leq X \leq 1.005$ they become positive and dominant and for $1.1 \leq X \leq 2.66$ they again are negative decreasing to very small values.

The ratio $\dfrac{|\rho_{DE}|}{\rho_m}$ is growing from $5 \times 10^{-33}$ at $X = 0.01$, up to 2.48 at $X = 1$ and then it decreases to zero at $X_{\max} = 2.66$.

Finally, the acceleration parameter $Q \equiv \dfrac{\ddot{a}}{a} \dfrac{1}{h^2}$ is calculated according to (47a). At $X = 0.01$ one has $Q = -0.5$. $Q$ remains negative up to $X \approx 0.9$ and becomes positive in the interval $0.995 \leq X \leq 1.005$; at $X = 1$ there is $Q = 0.5$. From $X = 1.1$ and up to $X_{max} = 2.66$, $Q$ is negative achieving $Q = -3$ at $X_{max} = 2.66$.

Table 1. Dynamics in the dust dominated period.

| $X \equiv \dfrac{a}{a_N}$ | $h$ (cm$^{-1}$) | $Q$ | $\rho_m$ (cm$^{-2}$) | $\rho_{DE}$ (cm$^{-2}$) | $P_{DE}$ (cm$^{-2}$) |
|---|---|---|---|---|---|
| 0.01 | $4.1 \times 10^{-26}$ | $-5.0 \times 10^{-1}$ | $2.1 \times 10^{-52}$ | $-1.0 \times 10^{-84}$ | $4.0 \times 10^{-79}$ |
| 0.10 | $1.3 \times 10^{-27}$ | $-5.3 \times 10^{-1}$ | $2.1 \times 10^{-55}$ | $-2.1 \times 10^{-83}$ | $1.5 \times 10^{-80}$ |
| 0.2 | $4.4 \times 10^{-28}$ | $-5.6 \times 10^{-1}$ | $2.6 \times 10^{-56}$ | $-6.4 \times 10^{-83}$ | $3.2 \times 10^{-81}$ |
| 0.5 | $1.4 \times 10^{-28}$ | $-3.4 \times 10^{-1}$ | $1.7 \times 10^{-57}$ | $-1.24 \times 10^{-77}$ | $8.0 \times 10^{-76}$ |
| 0.8 | $3.6 \times 10^{-29}$ | $-3.1 \times 10^{-1}$ | $4.0 \times 10^{-58}$ | $-1.5 \times 10^{-67}$ | $4.7 \times 10^{-66}$ |
| 0.9 | $6.0 \times 10^{-29}$ | $-3.0 \times 10^{-1}$ | $2.8 \times 10^{-58}$ | $-1.3 \times 10^{-62}$ | $5.4 \times 10^{-61}$ |
| 0.995 | $7 \times 10^{-29}$ | $+4.7 \times 10^{-1}$ | $2.2 \times 10^{-58}$ | $+3.6 \times 10^{-58}$ | $-2 \times 10^{-60}$ |
| 1.0 | $7.2 \times 10^{-29}$ | $+5.0 \times 10^{-1}$ | $2.1 \times 10^{-58}$ | $+5.2 \times 10^{-58}$ | $-2.6 \times 10^{-57}$ |
| 1.005 | $6.7 \times 10^{-29}$ | $0.1 \times 10^{-5}$ | $1.9 \times 10^{-58}$ | $2.7 \times 10^{-58}$ | $-1.8 \times 10^{-60}$ |
| 1.1 | $3.0 \times 10^{-29}$ | $-9.0 \times 10^{-1}$ | $1.5 \times 10^{-58}$ | $-5.0 \times 10^{-60}$ | $-5.0 \times 10^{-62}$ |
| 1.5 | $1.2 \times 10^{-29}$ | $-1.33$ | $6.1 \times 10^{-59}$ | $-1.4 \times 10^{-72}$ | $-1.6 \times 10^{-73}$ |
| 2.0 | $6.6 \times 10^{-30}$ | $-2.0$ | $2.5 \times 10^{-59}$ | $-1.3 \times 10^{-92}$ | $-4.5 \times 10^{-93}$ |
| 2.66 | $4.8 \times 10^{-30}$ | $-3.0$ | $2.0 \times 10^{-59}$ | $0$ | $0$ |

**7. A massive body in the homogeneous universe.**

In the previous sections a closed homogeneous isotropic universe filled with ordinary matter as well with Weylian DM and DE was considered. The line element (15) for that homogeneous universe was written as

$$ds^2 = d\bar{t}^2 - a^2(t)\left[\frac{dr^2}{1-r^2} + r^2 d\vartheta^2 + r^2 \sin^2 \vartheta d\varphi^2\right] \qquad (15)$$

For the forthcoming discussion it is convenient to write the line-element (15) in a slightly different form. With the very simple transformation, $dt = a(\tau) \cdot d\tau$, we get the conformal line element

$$ds^2 = a^2 d\tau^2 - a^2 \left(\frac{dr^2}{1-r^2} + r^2 d\vartheta^2 + r^2 \sin^2 \vartheta d\varphi^2\right) \qquad (15a)$$

Let us assume a body having mass **m** is located in any point of the dust dominated universe. Then, in addition to the time-dependent cosmic field discussed in the previous sections there is a spherically symmetric gravitational field surrounding the mass. This is the case to be considered in the present section. It will be described by the line-element

$$ds^2 = a^2 e^\nu d\tau^2 - a^2\left(\frac{e^\lambda dr^2}{1-r^2} + r^2 d\vartheta^2 + r^2 \sin^2\vartheta\, d\varphi^2\right) \qquad (55)$$

with $a(\tau), \lambda(r)$ and $\nu(r)$.

In the following will appear quantities depending on both, $r$ and $\tau$; which will be marked by a tilde, e. g. $\tilde{f}(\tau,r)$; a over-bar is used to mark quantities depending on $\tau$ solely, e. g. $\bar{f}(\tau)$. Finally, quantities depending on $r$ solely are not marked, e. g. $f(r)$. The universe is filled with conventionally matter, Weylian DM and Weylian DE, the latter being represented by $\tilde{\beta}$. These substances are now described by mass density $\tilde{\rho}(\tau,r)$, pressure $\tilde{P}(\tau,r)$ and Dirac's gauge function $\tilde{\beta}(\tau,r)$ all being functions of $\tau$ and $r$. The global field equation (14), with $\beta$ replaced by $\tilde{\beta}(\tau,r)$ and $T_\mu^\nu$ by $\tilde{T}_\mu^\nu$ is now written as

$$G_\mu^\nu = -\frac{8\pi}{\tilde{\beta}^2} \tilde{T}_\mu^\nu{}_{\text{matter}} + \frac{1}{\beta^2}\left(4\tilde{\beta}_\mu \tilde{\beta}^\nu - \delta_\mu^\nu \tilde{\beta}^\lambda \tilde{\beta}_\lambda\right) + \frac{2}{\beta}\left(\delta_\mu^\nu \tilde{\beta}_{;\lambda}^\lambda - \tilde{\beta}_{;\lambda}^\nu \delta_\mu^\lambda\right) - \delta_\mu^\nu \tilde{\beta}^2 \Lambda \qquad (56)$$

In (55) there are time/space separations of the metric coefficients e. g. $g_{00} = a(\tau)\cdot e^{\nu(r)}$. In accordance with this we assume $\tilde{\beta}(\tau,r) \equiv \bar{\beta}(\tau)\cdot \beta(r)$; where $\bar{\beta}(\tau)$ is the Dirac gauge function in the total homogeneous universe and $\beta(r)$ is caused by the massive body. It is also convenient to consider modified gauge functions $\tilde{b} \equiv \ln\tilde{\beta} \equiv \ln\bar{\beta} + \ln\beta = \bar{b}(\tau) + b(r)$, so that denoting $\dot{f} \equiv \frac{\partial f}{\partial \tau}$ and $f' \equiv \frac{\partial f}{\partial r}$ we have

$$\frac{\beta'}{\beta} = b'; \quad \frac{\dot{\bar{\beta}}}{\bar{\beta}} = \dot{\bar{b}}; \quad \frac{\beta''}{\beta} = b'' + (b')^2; \quad \frac{\ddot{\bar{\beta}}}{\bar{\beta}} = \ddot{\bar{b}} + (\dot{\bar{b}})^2; \quad \frac{\dot{\tilde{\beta}}'}{\tilde{\beta}} = \dot{\bar{b}} b'.$$

Finally, from EQ-s (55) and (56) one can obtain the non-zero gravitational $G_{0;0}^0; G_1^0, G_1^1; G_2^2$- equations explicitly. In these EQ-s we will represent components of the energy-momentum tensor as products of $\tau$ – depending factors and *r*-depending ones, $\tilde{T}_\nu^\mu(\tau,r) = \bar{T}_\nu^\mu(\tau)\cdot T_\nu^\mu(r)$, where $\bar{T}_\nu^\mu(\tau)$ is that of the total homogeneous universe.

The explicitly written equations are:

$$G_0^0: -3\frac{e^{-\nu}}{a^2}\left(\frac{\dot{a}}{a}\right)^2 + (1-r^2)\frac{e^{-\lambda}}{a^2}\left[\frac{\nu'-\lambda'}{r} - \frac{\nu'}{r} - \frac{2}{(1-r^2)} + \frac{1}{r^2}\right] - \frac{1}{a^2 r^2} = -\frac{8\pi}{\tilde{\beta}^2}\tilde{T}_0^0 +$$
$$+ 3\frac{e^{-\nu}}{a^2}\left[(\dot{\bar{b}})^2 + 2\left(\frac{\dot{a}}{a}\right)\dot{\bar{b}}\right] + (1-r^2)\frac{e^{-\lambda}}{a^2}\left[\lambda' b' + \frac{2rb'}{(1-r^2)} - \frac{4b'}{r} - 2b'' - (b')^2\right] - \tilde{\beta}^2 \Lambda \qquad (57a)$$

$$G_1^1: \frac{e^{-\nu}}{a^2}\left[\left(\frac{\dot{a}}{a}\right)^2 - 2\frac{\ddot{a}}{a}\right] + (1-r^2)\frac{e^{-\lambda}}{a^2}\left[\frac{\nu'}{r} + \frac{1}{r^2}\right] - \frac{1}{a^2 r^2} = -\frac{8\pi}{\tilde{\beta}^2}\tilde{T}_1^1 +$$
$$+ \frac{e^{-\nu}}{a^2}\left[2\ddot{\bar{b}} + (\dot{\bar{b}})^2 + 2\left(\frac{\dot{a}}{a}\right)\dot{\bar{b}}\right] - \frac{(1-r^2)\cdot e^{-\lambda}}{a^2}\left[3(b')^2 + \nu' b' + \frac{4b'}{r}\right] - \tilde{\beta}^2 \Lambda \qquad (57b)$$

and

$$G_1^0: \quad \frac{e^{-\nu}}{a^2}\left[\left(\frac{\dot{a}}{a}+\dot{\bar{b}}\right)(\nu'+2b')\right] = \frac{8\pi}{\tilde{\beta}^2}\tilde{T}_1^0 \equiv \frac{8\pi}{\tilde{\beta}^2\bar{\beta}^2}\overline{T}_1^0 T_1^0 \qquad (57c)$$

In EQ. (57c), $\overline{T}_1^0(\tau)$ is the radial energy flow in the homogeneous universe. However, from homogeneity and isotropy of that model follows $\overline{T}_1^0 = 0$. Thus,

$$\frac{e^{-\nu}}{a^2}\left[\left(\frac{\dot{a}}{a}+\dot{\bar{b}}\right)(\nu'+2b')\right] = 0 \qquad (58)$$

From (58) one concludes that either $\frac{\dot{a}}{a}+\dot{\bar{b}}=0$ or $\nu'+2b'=0$. As the first condition would be imposed on the homogeneous universe, in which $\bar{b} \equiv \ln\bar{\beta}$ was chosen according to cosmological observations, we accept the second condition.

Thus, the following gauging condition is accepted hereafter

$$b' = -\frac{1}{2}\nu' \qquad (59)$$

Making use of (59) we obtain the following gravitational equations.

$$G_0^0: \quad -3\frac{e^{-\nu}}{a^2}\left(\frac{\dot{a}}{a}+\dot{\bar{b}}\right)^2 + \frac{(1-r^2)\cdot e^{-\lambda}}{a^2}\left[-\nu''-\frac{3\nu'}{r}+\frac{\nu'-\lambda'}{r}+\frac{\lambda'\nu'}{2}+\right.$$
$$\left.+\frac{r\nu'}{1-r^2}+\frac{(\nu')^2}{4}-\frac{2}{1-r^2}+\frac{1}{r^2}\right] - \frac{1}{a^2r^2} = -\frac{8\pi}{\tilde{\beta}^2}\tilde{T}_0^0 - \tilde{\beta}^2\Lambda \qquad (60)$$

$$G_1^1: \quad \frac{e^{-\nu}}{a^2}\left[\left(\frac{\dot{a}}{a}-\dot{\bar{b}}\right)^2 - 2\left(\frac{\ddot{a}}{a}+\ddot{\bar{b}}\right) - 2(\dot{\bar{b}})^2\right] + \frac{(1-r^2)\cdot e^{-\lambda}}{a^2}\left[\frac{1}{r^2}-\frac{\nu'}{r}+\frac{(\nu')^2}{4}\right] -$$
$$-\frac{1}{a^2r^2} = -\frac{8\pi}{\tilde{\beta}^2}\tilde{T}_1^1 - \tilde{\beta}^2\Lambda \qquad (61)$$

$$G_2^2; \quad \frac{e^{-\nu}}{a^2}\left[\left(\frac{\dot{a}}{a}-\dot{\bar{b}}\right)^2 - 2\left(\frac{\ddot{a}}{a}+\ddot{\bar{b}}\right) - 2(\dot{\bar{b}})^2\right] +$$
$$+(1-r^2)\frac{e^{-\lambda}}{a^2}\left[\frac{\lambda'\nu'}{4}+\frac{\nu'-\lambda'}{2r}-\frac{\nu''}{2}+\frac{(3r^2-2)\nu'}{2r(1-r^2)}-\frac{1}{1-r^2}\right] = -\frac{8\pi}{\tilde{\beta}^2}\tilde{T}_2^2 - \tilde{\beta}^2\Lambda \qquad (62)$$

We recall that the time depended functions (the over-bared) were obtained in the homogeneous model. Thus, in (60-62) there are 3 equations for 5 functions $\lambda, \nu, T_0^0, T_1^1, T_2^2$.

From (61, 62) we can write

$$\frac{\nu''}{2}+\frac{(\nu')^2}{4}-\frac{\lambda'\nu'}{4}+\frac{\lambda'-\nu'}{2r}-\frac{r\cdot\nu'}{2(1-r^2)}+\frac{1}{r^2(1-r^2)}-\frac{e^\lambda}{r^2(1-r^2)} = \frac{8\pi\cdot a^2}{\tilde{\beta}^2(1-r^2)}(\tilde{T}_2^2-\tilde{T}_1^1) \quad (63)$$

This may be simplified by the following assumption. With $\tilde{T}_1^0 = 0$, no radial flow of matter is present and we can consider the substance in a system of reference where it is resting, so that $\tilde{T}_2^2 = \tilde{T}_1^1$, (this is of course holding in the homogeneous universe). By this assumption EQ. (63) may be written as

$$\frac{v''}{2} + \frac{(v')^2}{4} - \frac{\lambda' v'}{4} + \frac{\lambda' - v'}{2r} - \frac{r \cdot v'}{2(1-r^2)} + \frac{1}{r^2(1-r^2)} - \frac{e^\lambda}{r^2(1-r^2)} = 0 \qquad (64)$$

In (64) appear the metric functions $\lambda$ and $v$. According to the line element (55) let us assume

$$\frac{e^{\lambda+v}}{1-r^2} = C, \text{ with } C = Const > 0. \qquad (65)$$

Inserting this into (64) we get

$$v'' + (v')^2 - \frac{2v'}{r} + \frac{2}{r^2} - \frac{2Ce^{-v}}{r^2} = 0; \qquad (66)$$

and turning to $y \equiv e^v$ one obtains

$$y'' - \frac{2}{r} y' + \frac{2}{r^2} y = \frac{2C}{r^2}. \qquad (66a)$$

From which follows

$$y \equiv e^v = C + Ar + Br^2, \quad \text{and} \quad e^\lambda = \frac{C(1-r^2)}{C + Ar + Br^2} \qquad (67)$$

with $A$ and $B$ being arbitrary constants and $C > 0$ introduced in (65). We have also

$$\beta(r) \equiv e^b = \frac{\sqrt{C}}{\sqrt{C + Ar + Br^2}}, \qquad (67a)$$

One can consider a possible interpretation of solution (67). Denoting $\frac{A}{B} = -2m$; and $\frac{C}{B} = \kappa^2$ one rewrites $y \equiv e^v = C + Ar + Br^2$ as

$$y \equiv e^v = \frac{C}{\kappa^2} r^2 \left(1 - \frac{2m}{r} + \frac{\kappa^2}{r^2}\right); \qquad (68)$$

We will adopt the following meanings: **m** stands for the mass of the body and $\kappa$ characterizes the Weyl scalar charge that is created by the $\beta$ − DE around the mass **m**. It will be assumed that $\kappa^2 < m^2$, so, one can write $\kappa = nm$ with $n < 1$.
In (68) one recognizes (in the bracket) an a' la Reissner-Nordstrøm metric.
With $\kappa^2 < m^2$ and $C > 0$ one rewrites (68) as

$$y \equiv e^v = \frac{C}{\kappa^2}(r^2 - 2mr + \kappa^2) \qquad (69)$$

Here, as in the Reissner-Nordstrøm case, there are two horizons: the inner, Chauchi, horizon $r_{in}$ and the outer event one - $r_{out}$.

$$r_{in} = m - \sqrt{m^2 - \kappa^2} = m\left(1 - \sqrt{1-n^2}\right); \quad r_{out} = m + \sqrt{m^2 - \kappa^2} = m\left(1 + \sqrt{1-n^2}\right) \qquad (70)$$

Space and time change roles upon crossing the outer horizon, $r_{out}$, and change roles again upon crossing $r_{in}$.

## 8. Dark energy wrapping the massive body.

Dark energy represented by Dirac's gauge function $\beta$ does not interact with matter; it is affected only by geometry-gravitation. In the spherically symmetric gravitational field of the mass **m**, the $\beta$ – DE will form a ball-like concentration around **m**, -- DE wraps up the massive body.

In order to evaluate the mass of this $\beta$ – ball let us consider EQ. (57a). In absent of *r*-depending $\beta$ – DE the second brackets in the RHS vanishes, but with nonzero $b'; b''$ it may be regarded as the *r*-depending part of the mass-energy density of the $\beta$ – DE field around the mass **m**

$$-8\pi\rho_\beta = \frac{(1-r^2)e^{-\lambda}}{a^2}\left[\lambda' b' + \frac{2rb'}{(1-r^2)} - \frac{4b'}{r} - 2b'' - (b')^2\right] \qquad (71)$$

Inserting into (71) the values of $\lambda'; b'; b''$ (Cf. 67, 67a, 68) one obtains

$$8\pi\rho_\beta = \frac{1}{a^2\kappa^2}\left[\frac{4m}{r} - 6 + 3\cdot\frac{(r-m)^2}{(\kappa^2 - 2mr + r^2)}\right]. \qquad (72)$$

The mass of the DE ball at *r* may be accounted by

$$M_\beta = \int_{r_0}^{r}\rho_\beta dV = \int_{r_0}^{r}4\pi\rho_\beta\sqrt{C}\,\frac{r^2 dr}{\sqrt{y}} = \int_{r_0}^{r}\left[\frac{2m}{\kappa^2 r} - \frac{3}{\kappa^2} + \frac{3}{2\kappa^2}\frac{(r-m)^2}{(\kappa^2 - 2mr + r^2)}\right]\frac{a^{-2}\kappa r^2 dr}{\sqrt{r^2 - 2mr + \kappa^2}}$$

(73)

Let us consider the region outside the event horizon for $r_{out} < r < 1$. Integrating (73) one obtains (Cf. (Ryzhik & Gradshteyn, 1951))

$$M_\beta = \left|\frac{3r(n^2 m^2 - r^2) + 5m[r^2 - 2mr + n^2 m^2]}{4nma^2\sqrt{r^2 - 2mr + n^2 m^2}} + \frac{m(5-3n^2)}{4na^2}\cdot\ln\left[2\sqrt{r^2 - 2mr + n^2 m^2} + 2r - 2m\right]\right|_{r_0}^{r}$$

(74)

For $r_0$ we will take a value close to $r_{out} = m + \sqrt{m^2 - \kappa^2}$. Let us set $r_0 = 2m$. Then

$$m_\beta(r_0 = 2m) = \frac{m(11n^2 - 24)}{4n^2 a^2} + \frac{m}{4na^2}(5 - 3n^2)\ln[2m(1+n)] \qquad (75)$$

From (75) we have $M_\beta(r_0 = 2m) < 0$ and for $r > 2r_0$ the mass $M_\beta = M_\beta(r) - M_\beta(r_0)$ is **negative** too. For great enough **r**-s ($2m \ll r \leq 1$) with $n = 0.5$ one can write

$$M_\beta \approx \frac{1}{a^2} \left[ -\frac{3r^2}{2m} + \ln\left(\frac{4r}{3m}\right)^{2.125m} \right] \tag{76}$$

Taking into account that $m \ll 1$, one sees that the $\beta$ – DE ball has a negative mass.

Negative mass for large **r**-s may be understood if we go back to the energy density (72). For large enough $r - s$, i. e. $r \gg m$ we get a very simple result, $8\pi\rho_\beta \approx -\frac{3}{a^2\kappa^2}$. Thus, the concentration of $\beta$ – DE around the perturbing mass creates a DE-ball with negative energy-mass density and negative gravitational mass. The negative $M_\beta$ is universally **repulsive**, both positive-mass and negative-mass objects will be pushed away.

The repulsive force must be considered as a factor causing and supporting the acceleration of the expanding universe.

DE will also cause pressure. Let us consider the $G_1^1$ –equation (57b). The **r**-depending part of pressure created by the $\beta$ – DE is

$$8\pi P_{\beta(r)} = \frac{(1-r^2) \cdot e^{-\lambda}}{a^2} \left[ 3(b')^2 + v'b' + \frac{4b'}{r} \right] \tag{77}$$

Inserting $b'; v'; e^{-\lambda}$ we get

$$8\pi P_{\beta(r)} = \frac{1}{a^2\kappa^2} \left[ \frac{(r-m)^2}{r^2 - 2mr + \kappa^2} - \frac{4(r-m)}{r} \right] \tag{77a}$$

and for $m \ll r \leq 1$ the radial pressure as well the energy density (72) is

$$8\pi P_\beta \doteq 8\pi\rho_\beta \doteq -\frac{3}{a^2\kappa^2} \tag{78}$$

Let us take into account that the mass of a galaxy $m \ll 1$, so that in (76) the term $\ln\left(\frac{4r}{3m}\right)^{2.125m}$ is negligible. Thus, near the horizon of the universe the $\beta$ – DE ball has a negative mass $M_\beta \approx -\frac{3}{2ma^2}$ as well negative mass density and negative pressure.

## 9. Outlook and discussion

The considered cosmology is based on the Weyl-Dirac (W-D) theory that provides us with three geometrically based functions: the metric tensor $g_{\mu\nu} = g_{\nu\mu}$, the Weyl length connection vector $w_\nu$, the Dirac gauge function $\beta$. Instead of setting in the action integral for Dirac's (Dirac, 1973) parameter $\sigma = 0$, we take as proposed by Rosen (Rosen, 1982) $\sigma = -2\kappa^2$. By this choice we get dark matter particles.

On large scales the universe **is** homogeneous and isotropic; however it has a chaotic Weylian microstructure. In sections 4 and 6 the W-D spatially closed cyclic universe is considered.

The locally restricted chaotically oriented Weyl fields existing in micro-cells create bosons, (Cf. Sec. 5) named weylons, with mass $m_w = \dfrac{\kappa \hbar}{c}$ and spin 1. With $m_w > 10 \text{MeV}$ these particles form the cold dark matter bulk in the universe.

At the beginning of the expansion phase there was no matter; the embryo universe was a homogeneous, 3-dimensional spherically symmetric geometric entity having the radius $a_0 = 1.616 \times 10^{-33}$ cm and filled with Dirac's gauge function $\beta_{\text{create}}$. Very close to the beginning occurs the process of matter creation by $\beta_{\text{create}}$. Matter creation continues up to the moment, when the Planck density is achieved. From this on and up to $t_1 = 1.05 \times 10^{-31}$ cm, lasts the prematter period, characterized by the equation of state $P = -\rho$ with $\rho = \rho_{Pl}$. During this period there is an inflation-like growing of the radius, from $a_{Pl} = 1.2 \times 10^{-31}$ cm to $a_1 = 1.37 \times 10^{-5}$ cm and a growing of the temperature from $T_{Pl} = 2.7 \times 10^{-165} K$ to $T_1 = 5.24 \times 10^{17} K$.

After the prematter period the universe entered a transition period, during which prematter gradually transforms into radiation and ultrarelativistic particles. In the transition period, the temperature achieves its maximum $T_{\max} = 7.4 \times 10^{31} K$ and then decreases to $T_{\text{Rad}} = 2.43 \times 10^{27} K$. At these huge temperatures micro-cells of inhomogeneity appear, in which the local Weyl fields $w_{\nu_{\text{loc}}}$ create weylons. Following previous papers (Israelit & Rosen, 1992, 1994) we presume that the ensemble of weylons constitute a cold dark matter gas. As conventionally matter (particles and radiation) has no Weylian charge it will not interact with the weylon dark matter. The transition period is followed by the radiation period, which lasts from $a_{\text{Rad}} = 100$ cm to $a_{\text{Dust}} \approx 3.208 \times 10^{26}$ cm; during it the temperature of conventionally matter decreases from $T_{\text{Rad}} = 2.43 \times 10^{27} K$ to $T_{\text{Dust.conv.}} = 7.575 \times 10^2 K$, the weylons temperature is significantly lower. The radiation period turn into the dust dominated.

It is believed that during the early dust dominated epoch, close to $a_{\text{Dust}}$ and $T_{\text{Dust.conv.}}$, the galaxies were formed, so that plausibly the weylon DM, through its gravitational interaction with conventional matter, played an important spurring role in the galaxy formatting process.

In the dust period, the gauge function creates dark energy, which is insignificant in the early and late (beyond the present state) dust period, but become dominant at present, when it is causing cosmic acceleration.

After the present state the expansion continues up to $a_{max} = 2.66 a_N$, where one has $h^2 \approx 2.3 \times 10^{-59} \text{cm}^{-2}$ and a great deceleration parameter $Q \equiv \frac{\ddot{a}}{ah^2} = -3$. The latter will cause the universe to contract back to the beginning.

As said above, in sections 4 and 6 a W-D spatially closed cyclic cosmological model was considered. The space was taken as homogeneous and isotropic, filled with luminous and dark matter as well with dark energy. These ingredients were created by geometrically features of the W-D geometry and are described by functions depending on the cosmic time solely. The model is in agreement with cosmological observations and predicts the acceleration of the expanding universe at present.

However, the homogeneous universe may be considered only as the first, rough approximation, - the real universe contains stars, galaxies, clusters of galaxies etc together with DM and DE.

In section 7, a **massive body** (a star, galaxy, cluster) during the dust period in the homogeneous model is considered. This body will create a spherically symmetric gravitational field that is put together with the time dependent field of the homogeneous model. It is believed that from the field caused by a single massive body one can make conclusions about the influence of massive celestial bodies on the dynamics of the whole universe.

In the homogeneous universe, the Weyl DE penetrates the whole space uniformly. However, (Cf. section **7**) in the spherically symmetric gravitational field of a massive body, the $\beta$ – DE will form a ball-like concentration around the center of this body. It was found that this concentration of DE has a negative mass density, a negative pressure and a negative mass.

Negative mass in General relativity were considered by Hermann Bondi (Bondi, 1957) and by W. B. Bonnor (Bonor, 1989). The possibility of an elementary particle (electron?) containing negative mass was discussed by (Bonnor & Cooperstock, 1989). (Bondi, 1957) has considered an interesting problem of two bodies having equal absolute values of mass, but one a positive, the second a negative mass. In absent of other forces, this system would move with constant acceleration along the line of centers.

In order to consider the effect of cosmic acceleration by DE let as consider the situation near the horizon of the universe, $r = 1$. Here one has $8\pi \rho_\beta \doteq 8\pi P_\beta \doteq -\frac{3}{a^2 \kappa^2}$; and $M_\beta \approx -\frac{3}{2ma^2}$. The negative pressure is necessary for causing an acceleration of the expanding universe and a deceleration during the contracting phase. The equation of state of our DE indicates that near the horizon DE is in the state of stiff matter.

The negative $m_\beta$ is universally *repulsive*, both positive-mass and negative-mass objects placed near the horizon will be pushed away, by constant acceleration. The repulsive force is a factor causing and supporting the acceleration of the expanding universe. It is believed that negative values of $M_\beta$, $\rho_\beta$ and $P_\beta$ generally characterize DE.

According to the results of the present model there is no big bang; after achieving the maximum radius the universe is returning during a contraction phase to the initial

Planck state. Conventional matter, DM and DE are created by geometrically based functions.

*The Weyl-Dirac theory, which is a minimal expansion of Einstein's theory, is a geometrically based framework appropriate for describing and searching cosmology.*